\documentclass[11pt,preprint]{aastex}
\usepackage{epsfig}
\usepackage{color}
\def\new#1{\color{black}{#1}\color{black}}

\newcommand{\etal}{et~al. }

\newcommand{\msun}{{\rm M}_\odot}
\newcommand{\yr}{\,{\rm yr}}
\newcommand{\pyr}{\,\yr^{-1}}

\begin{document}
\title {Missing Iron Problem and Type Ia Supernova Enrichment of Hot Gas in Galactic 
Spheroids}
\author{\vspace{-\parskip}
Shikui Tang and Q. Daniel Wang}
\affil{\vspace{-\parskip}
  Department of Astronomy, University of Massachusetts,
  Amherst, MA 01003; tangsk@astro.umass.edu and wqd@astro.umass.edu  }

\begin{abstract}

Type Ia supernovae (Ia SNe) provide a rich source of iron for hot gas
in galactic stellar spheroids. However, the expected super-solar 
iron abundance of the hot gas is not observed. Instead, X-ray observations 
often show decreasing iron abundance toward galactic central regions,
where the Ia SN enrichment is expected to be the highest. 
We examine the cause of this missing iron problem by studying
the enrichment process and its effect 
on X-ray abundance measurements of the hot gas. The evolution of Ia SN
iron ejecta is simulated in the context of galaxy-wide hot gas outflows, in both
supersonic and subsonic cases, as may be expected for hot gas in 
galactic bulges or elliptical galaxies of intermediate masses.
SN reverse-shock heated iron ejecta is typically found to have  
a very high temperature and low density, hence producing
little X-ray emission. Such hot ejecta, driven by its large buoyancy,
can quickly reach a substantially higher outward velocity 
than the ambient medium, which is
dominated by mass loss from evolved stars. The ejecta is gradually and dynamically mixed with the medium at large galactic radii.
The ejecta is also slowly diluted and cooled by {\sl insitu} mass injection 
from evolved stars. These processes together naturally result in the observed
positive gradient in the average radial iron abundance 
distribution of the hot gas, even if mass-weighted.
This trend is in addition to the X-ray measurement bias that
tends to underestimate the iron abundance for the hot gas with 
a temperature distribution.
\end{abstract}

\section{Introduction}
Soft X-ray emission of hot gas in and around galaxies depends heavily 
on the properties of metals (i.e., elements heavier than helium). 
Due to the concentrated enrichment of Type Ia supernovae
(Ia SNe), the primary iron provider of the universe, hot gas in 
galactic stellar spheroids (galactic bulges and elliptical
galaxies), in particular, is expected to have a super-solar iron abundance.
Away from such a stellar spheroid, the iron-rich hot gas
might be diluted by in-falling low-abundance gas.
Thus a monotonically decreasing iron abundance profile
as a function of galactic radius is also expected
(e.g., \citealt{Buote00apj,Borgani08}) and is indeed
observed on scales of the intragroup and intracluster media.

On smaller scales (within individual stellar spheroids and their immediate 
vicinity), however, X-ray-inferred iron abundances of the hot 
gas seem to be at odds with the predictions. 
The abundances are typically sub-solar for low- and intermediate-mass 
spheroids with log($L_x) \lesssim 41$ to about solar for more massive ones 
(e.g., \citealt{Humphrey06}), substantially less than what
are expected from the Ia SN enrichment.
Furthermore, a number of galaxies show a significant
iron abundance drop toward their central regions, where the stellar 
feedback should be the strongest. Examples of this abundance drop 
include M87 \citep{gm02}, 
NGC 4472, NGC 5846 \citep{Buote00apj}, and NGC 5044 \citep{buo03}. 
The inward drop of the iron abundance and its low value in general
are so far not understood, which may indicate that only a small 
portion of metals produced by Ia SNe is observed, while the rest is 
either expelled or in a state that the present X-ray data are not sensitive to. 

\citet{Brighenti2005} have explored the possibility that
the iron produced by Ia SNe may have radiatively
cooled and may then have avoided the detection. Based on
a 1-D modeling of Ia SN remnant (SNR) evolution,
they have demonstrated that the iron-rich ejecta may cool rapidly to low
temperatures because of its large radiative emissivity.
This scenario assumes that the cooling occurs within
a critical time scale for the ejecta to be mixed microscopically
with the ambient medium to an iron abundance below about 100 solar. 
This critical time scale is not determined, but should partly depend on
the uncertain magnetic field strength and configuration, which 
can substantially affect the effectiveness of the iron diffusion.
The mixing, of course, can also be dynamic in the violent environment
of galactic stellar spheroids. 

We have studied the Ia SNR evolution in a more realist
setting. We have conducted 3-D hydrodynamic simulations of hot gas in
a typical galactic stellar spheroid of an intermediate-mass (i.e., a 
galactic halo mass $M_h \sim 10^{12} M_\odot$) in
a relatively isolated environment. The
hot gas is expected to be in a galaxy-wide outflow, either supersonic or 
subsonic, depending largely on the feedback history of 
the galaxy \citep{Tang09a}. \citet{Tang09b} have presented simulations
for the supersonic outflow case and have discussed several important 
3-D effects:
(1) SN ejecta is not well mixed with the ambient medium, and soft X-ray 
emission arises primarily from relatively low temperature 
gas shells associated with SN blastwaves; (2)  The inhomogeneity in the 
gas temperature substantially alters the emission spectral shape, 
which can lead to an artificially lower iron abundance (by a factor of 2-3) 
in a spectral fit with a simplistic thermal plasma model; (3) 
The average 1-D and 3-D simulations give substantially different radial 
temperature profiles (e.g., the 
inner temperature gradient in the 3-D simulation is positive, mimicking a 
``cooling flow''); (4) The inhomogeneity also enhances the diffuse 
X-ray luminosity by a factor of a few.

In the present work, we focus on the evolution of SN ejecta 
and its effect on the average iron abundance measurements of the hot gas 
in both supersonic and subsonic outflow cases. We use the 3-D simulations
to study the dilution of Ia SN iron ejecta by the stellar mass loss and the 
dynamic mixing with the ambient medium as well as the bulk radial outflows 
at abundance-dependent velocities. 
We demonstrate that the dropout of the iron ejecta is
unlikely to be important because of the effective dynamic mixing and
dilution. But the differential outflows and their gradual mixing
can naturally result in an iron abundance profile qualitatively similar to 
what is observed. The non-uniformity of the iron abundance in the flow 
could also lead to an underestimate of the globally averaged iron abundance
in X-ray spectral measurements. These effects are in 
addition to the bias that is caused by the broad temperature distribution of 
the hot gas considered in \citet{Tang09b}, where only supersonic outflows 
are simulated and a uniform solar abundance is assumed for mock X-ray 
spectral analysis. 

The rest of the paper is organized as follows: We first describe in 
\S~2 our model inputs and simulation setups and then present  in \S~3
our new results; We give simple explanations for a few key 
phenomena in \S~4 and summarize the work in \S~5.
Throughout of the paper, we adopt the solar metal abundances
as  listed in \citet{Anders89}; in particular, our quoted iron 
mass abundances (i.e., \new{the ratio of iron mass to hydrogen 
mass},) are relative to the solar value $2.6\times 10^{-3}$.

\section{Model Inputs and Simulation Setups}

\begin{deluxetable}{lr}
\tabletypesize{\footnotesize}
\tablecaption{Basic Model Parameters}
\tablewidth{0pt}
\tablehead{
Parameter & Value
}
\startdata
Present Mass loss rate ($\rm\msun\pyr$) & 0.075 \\
Present energy input rate ($\rm\, erg\,s^{-1}$) &$1.2\times 10^{40}$ \\
Bulge stellar mass $M_*$ (M$_\odot$) & $3 \times 10^{10}$ \\
Bulge scale $r_b$ (kpc) & 0.7 \\
Halo mass $M_h$ (M$_\odot$) & $10^{12}$ \\
\enddata
\label{tab:model}
\end{deluxetable}

\new{Ideally we should conduct the high-resolution 3-D simulations of the entire 
life of a galaxy. 3-D effects can be important in determining the early
evolution of a galactic spheroid, especially in its starburst stage when
the gas is in multiple phases \citep{Tang09b}. But at present such a
simulation is far too expensive to be performed with sufficient
resolution to trace the detailed chemical
enrichment. Thus, we have modeled the evolution of large-scale gaseous outflows
in 1-D simulations to capture the global hot gas dynamics around the galaxies
\citep{Tang09b}. In the present paper, we focus on  3-D 
effects of the Ia SN enrichement in the globally established hot gas 
outflows in a region of a few kpc radius 
around the center of a spheroid to facilitate potential direct 
comparison with X-ray observations. Across this region, the dynamic time scale
is short, allowing for detailed 3-D simulations. We find that the 3-D effects 
are important for small-scale density, temperature, and abundance structures, 
but have little impact on the global average dynamics of the outflows. 
Therefore, our approach is self-consistent.}

The present investigation is based primarily on two 3-D simulations of hot gas
in galaxy-wide outflows driven by Ia SNe in galactic stellar spheroids. The 
supersonic case corresponds to the bulk outflow that reaches a sonic point 
(at radius $\sim 1.4$ kpc), whereas the subsonic one does not.
The basic setup for these two cases
is the same; the model parameters are listed in Table~\ref{tab:model}. 
The only difference between the two is the setup of the initial conditions:
the supersonic case starts from the 1-D solution, similar to what is 
presented in \citet{Tang09b}, whereas the subsonic case starts from the final 
output of a 1-D simulation, which includes the evolution of 
the circum-galactic medium in a more realistic context of galaxy
evolution (Model VE; \citealt{Tang09a}). 
By comparing results from the two cases,
we intend to illustrate the dependence of the iron ejecta evolution
on the global outflow speed. 

Both the gas mass and the mechanical energy 
injections are assumed to follow the same smooth distribution as the stars,
which is a Hernquist profile \citep{Hernquist1990} with the characteristic 
radius of $r_b$. The uncertainties in the mass and 
energy injection rates are a factor of $\sim 2$, depending on specific
empirical calibrations, potential mass-loading from cool gas, etc. 
(e.g., \citealt{David06,Tang09a} and references therein). Our chosen energy 
rate is on the lower side of the uncertainty range, whereas the mass rate is
on the higher side. These choices result in on average a lower gas temperature,
a greater soft X-ray emissivity, and a lower iron abundance, which are more
close to what are directly inferred from X-ray observations. 
\new{In particular, existing studies have shown that the soft X-ray luminosities 
of spheroids are typically substantially higher than those predicted by the 
supersonic wind model, indicating that hot gas may most likely be in 
subsonic outflows \citet{Tang09a,Tang09b}. However, this conclusion is still 
very tentative; a more careful systematic analysis of existing data is 
required. For low- or intermediate-mass spheroids, the soft X-ray contribution 
from faint cataclysmic variables and coronally active binaries becomes 
important, which needs to be carefully subtracted. Indeed, the very faint
diffuse X-ray emission from NGC 3379 after the subtraction of this 
contribution seems to be consistent with the superwind model \citep{tri08}. 
Therefore, our simulations of both the supersonic
wind and the subsonic outflow should provide a qualitative characterization
of the 3-D effects on the Ia SN chemical enrichment of hot gas for 
both low- and intermediate-mass spheroids. }
The 1-D models also assume an instantaneous mixing between the ejecta 
and stellar mass injection and hence a uniform iron abundance 
of 2.7 solar. We have ignored any feedback from the supermassive
black hole expected to be present at the center of a spheroid. 
This feedback, likely occurring in bursts with certain preferential 
directions (e.g., in form of jets), can occasionally result in significant 
disturbances in global hot gas distributions, as reflected 
by the asymmetric X-ray morphologies observed in some elliptical 
galaxies \citep{die08}. But, averaged over the time, Ia SNe are 
energetically more important for a low- or intermediate-mass spheroid with 
$L_K \lesssim 10^{11}L_{\odot K}$ \citep{David06}.

The simulations are performed with FLASH \citep{Fryxell00}, 
an Eulerian astrophysical hydrodynamics code
with the adaptive mesh refinement (AMR) capability.
The 3-D simulated box, 128 kpc on a side, is centered on the spheroid 
and has the so-called outflow (sometimes called
zero-gradient) boundary condition. As in \citet{Tang09b}, only one 
octant of the box is simulated at full resolution
(down to about 4 pc), while the resolution is degraded by a factor of four
in the rest of the grid, except for regions where SNRs seeds have just 
been embedded. The difference in the resolution between the regions of 
the same simulation allows us to check 
the resolution effect (e.g., on the X-ray luminosity), which appears 
to be rather small. In the 3-D simulations, the Ia SN injection follows 
the star distribution only statistically \citep{Tang09b}. 
For each ``randomly'' generated SN, its remnant seed is extracted from a
library of 1-D simulated radial density, temperature, and 
velocity profiles and is scaled appropriately before being
planted into the 3-D simulations \citep{Tang05,tw09}. 

We evolve the two simulations till the averaged
radial profiles of the density, pressure, and abundance have reached 
a statistically steady state inside the galactic radius of 10 kpc. For 
any quantitative analysis, we will use only the data within this radius to
avoid any potential artifacts introduced by the chosen boundary condition,
in the high-resolution octant, and during the final stabled period. 

\begin{figure}[bthp]
\epsscale{0.94}
\begin{center}
\plotone{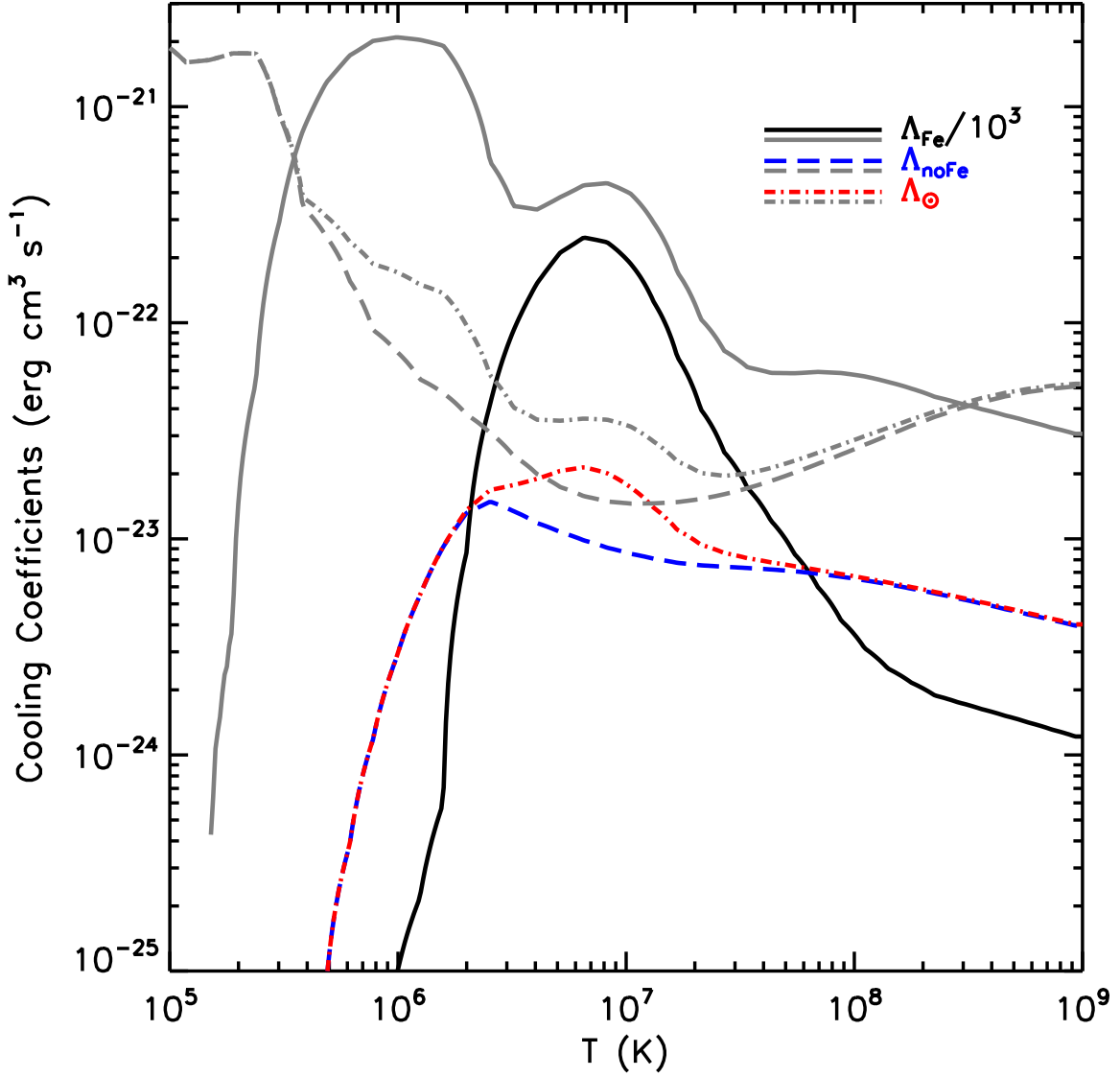}
\caption[The radiative coefficients]{\label{F:coolfuns}
Radiative coefficients of optically-thin thermal plasma 
with various abundance compositions: pure iron ($\Lambda_{\rm Fe}$,
which is decreased by a factor of $10^3$
for easy visualization;
solid lines), solar abundance without iron
($\Lambda_{\rm noFe}$; long-dash lines),
and solar abundance ($\Lambda_{\odot}$; dash-dot lines). All these
coefficients are constructed  from {\small XSPEC} using the {\small MEKAL}
model.
The black and colored lines denote the coefficients
in 0.3-2.0 keV band, while the corresponding gray lines are
in the bolometric band.
Note that $\Lambda_{\odot} = \Lambda_{\rm noFe}
+ 4.68\times 10^{-5} \Lambda_{\rm Fe}$.
}
\end{center}
\epsscale{1.0}
\end{figure}

For the calculation of the X-ray emission and the cooling of the hot gas,
we use the {\small MEKAL} plasma model \citep{Mewe85,Liedahl95},
extracted from the {\small XSPEC} package. The cooling function of
the gas is divided into the iron-free part
$\Lambda_{\rm no Fe}(T)$ and the pure iron rate $\Lambda_{_{\rm Fe}}(T)$,
which are shown in Fig.~\ref{F:coolfuns} (see also \citealt{Brighenti2005}).

\section{Results}

\begin{figure}[hbpt]
\begin{center}
\plotone{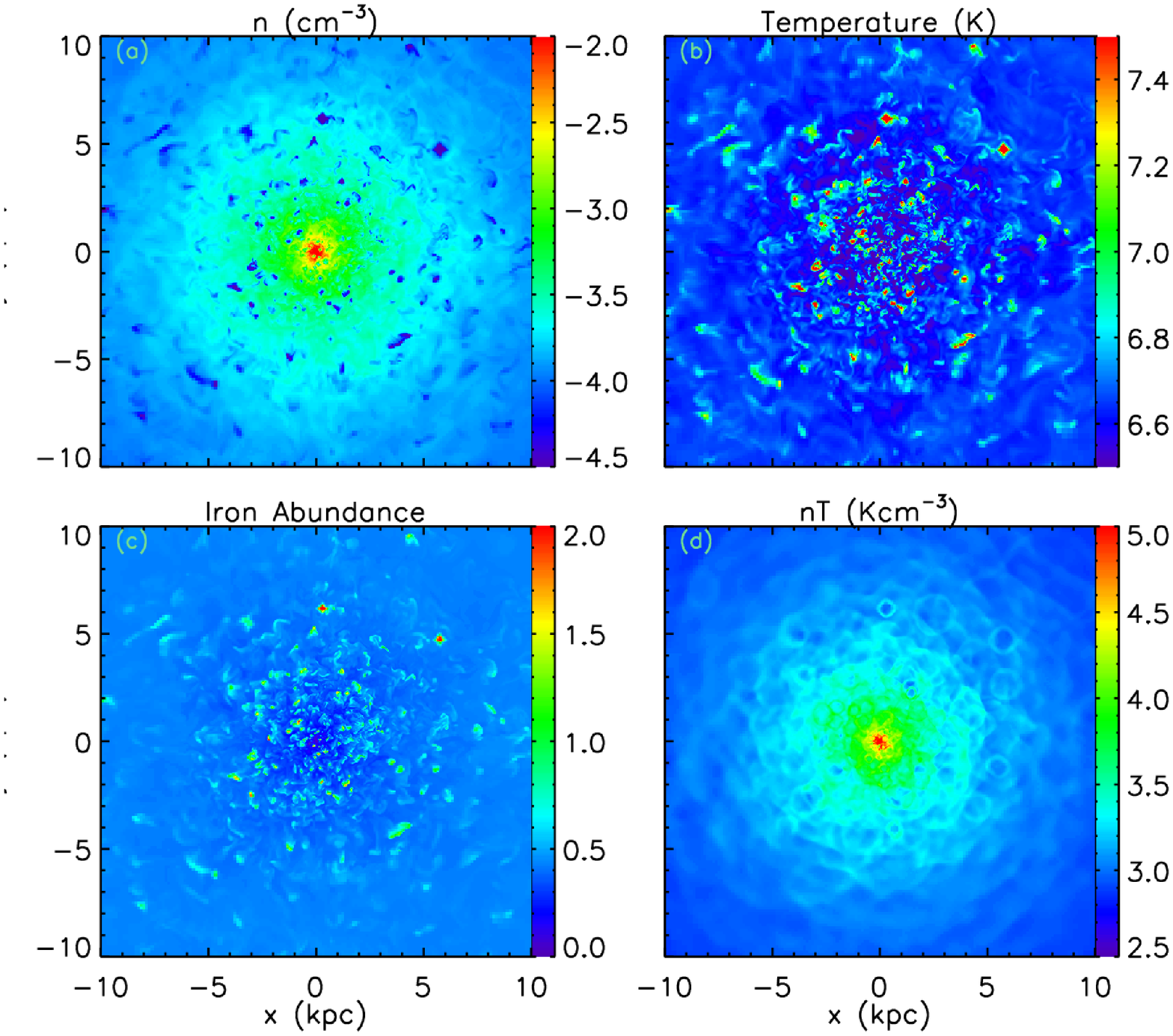}
\caption[]
{\label{F:global} Snapshot of the 3-D simulated subsonic outflow 
in the z = 8 pc plane, showing the density (a), temperature (b), iron 
abundance (c), 
and pressure of the gas (d). All plots are logarithmically
scaled according to the color bars. 
Note that the upper right quarter region in each
panel denotes the data from the octant at full resolution.
}
\end{center}
\end{figure}

\begin{figure}[hbpt]
\begin{center}
\plotone{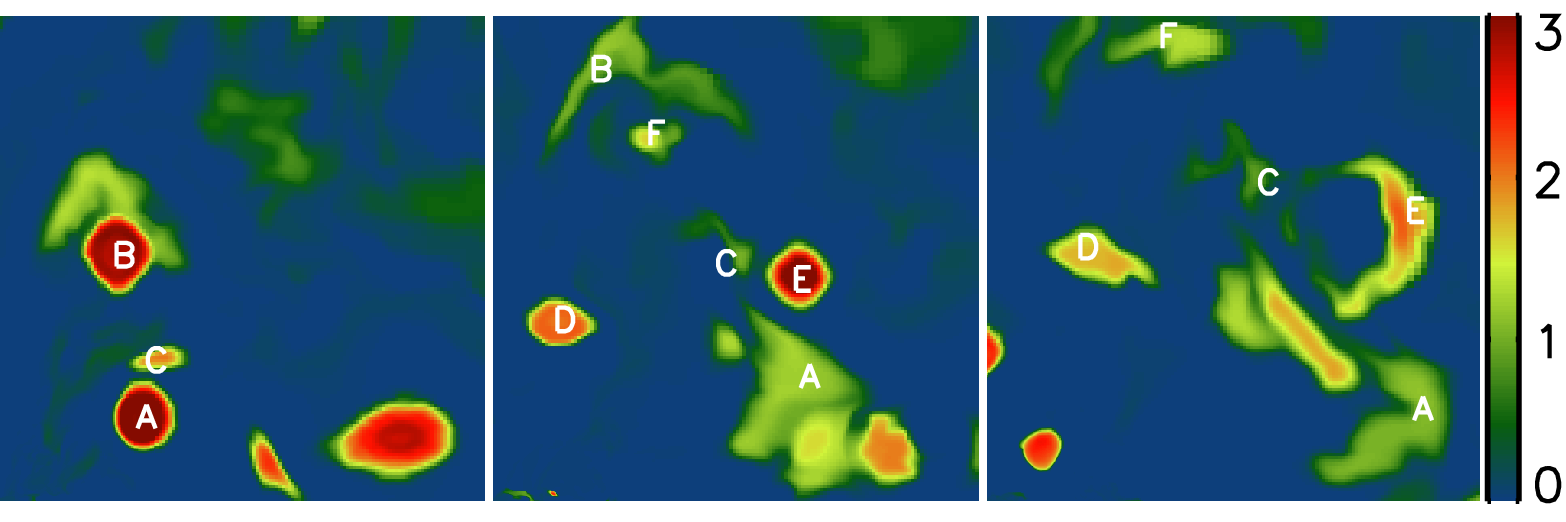}
\plotone{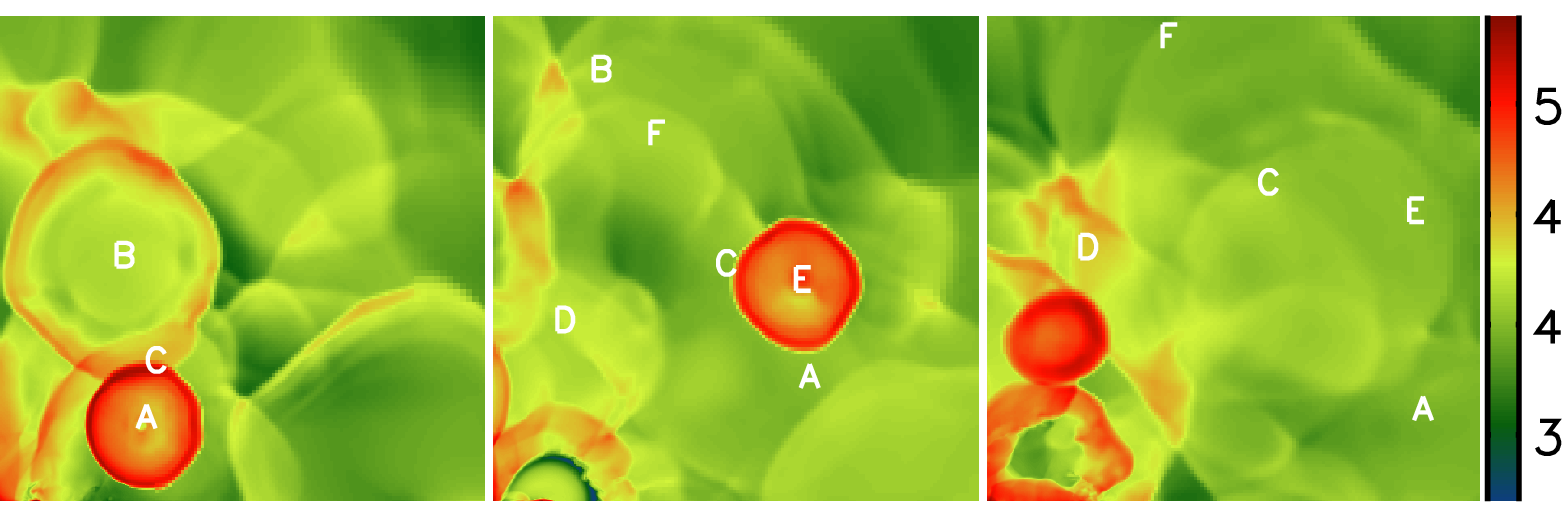}
\caption[The iron ejecta evolution of a few SNe]
{\label{F:ironejecta}
Close-ups of SNR evolution: iron abundance  mostly sensitive to SN ejecta
(upper row) and pressure
sensitive to outer blastwave shells (lower row). The color
scales are plotted logarithmically. 
The three panels in each row represent the same physical area 
of 0.6 kpc $\times$ 0.6 kpc; the low left corner of each panel is projected 
to the center of the model spheroid; the middle and right panels are snap-shots 
taken 0.86 and 1.52\,Myr later after
the left one. Features that orignated from the same SN ejectors are marked with the 
same letters in the panels. The apparent systematic motion of the features
to the right and/or top of the area, as seen from the left to right panels, is due to the 
bulk outward flow of the hot gas. The brightest circular red disk in the 
lower right panel represents a remnant that is cut on a side that does not 
cross the SN iron ejecta and thus does not show up in the upper 
right panel. 
}
\end{center}
\end{figure}

Fig.~\ref{F:global} shows a sample snapshot of our simulated subsonic 
outflow, with perspectives on the density,
temperature, pressure, and iron mass fraction distributions across one
plane cutting through the simulated box. The highly structured distributions
of the hot gas are similar to those seen in the supersonic case 
(\S~1; \citealt{Tang09b}), 
although features in the subsonic case tend to be more compact 
because of its higher average gas density. Time-sequenced
close-ups of a region near the center of the subsonic outflow is presented in 
Fig.~\ref{F:ironejecta}, 
demonstrating the evolution of several SNRs as well as the bulk motion. The feature labeled with ``A''
in the upper left panel, for example, represents the iron
core of a newly embedded SNR. As indicated in subsequent panels,
the iron core is gradually diluted by the local mass injection
from evolved stars. In the mean time, the shapes of the core as well as the 
outer shell are strongly affected by the impacts of nearby SNRs (lower row).
Because of both the dilution and the mixing with the 
surrounding medium, the iron abundance of the core is reduced to
about 10 solar after about 1.5\,Myr (right panel). This time scale is much too
short to allow for significant cooling of the ejecta.
\new{Due to the limited spatial resolution of the simulations, the resultant iron distribution should be slightly more diffusive than the reality. But the iron mixing appears to be dominated by the random impacts or gas turbulence. In fact, we find no significant difference between the iron distributions extracted from the low and high resolution octants.}

\epsscale{1.0}

\epsscale{1.0}

\begin{figure}[tbph]
\begin{center}
\plottwo{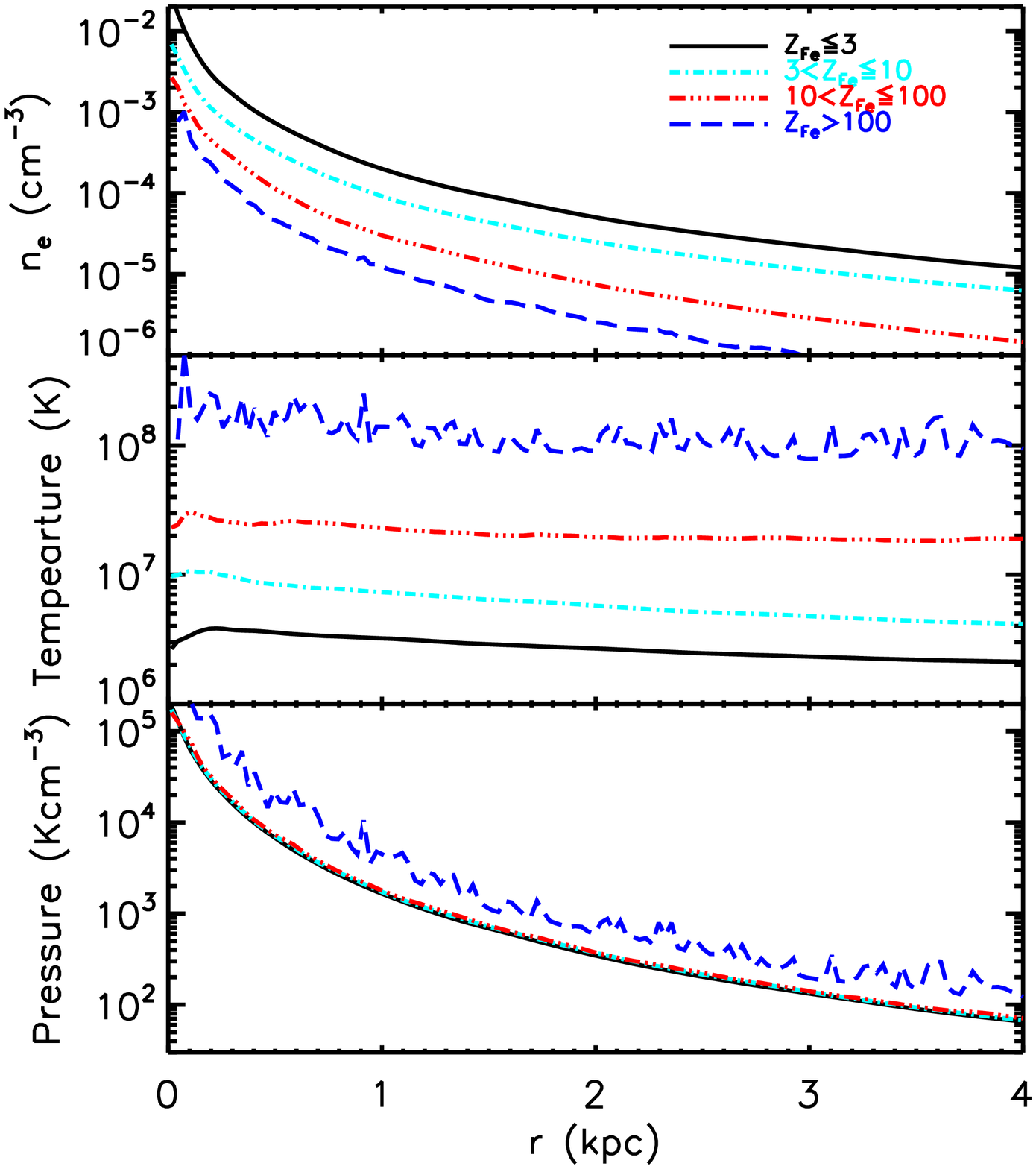}{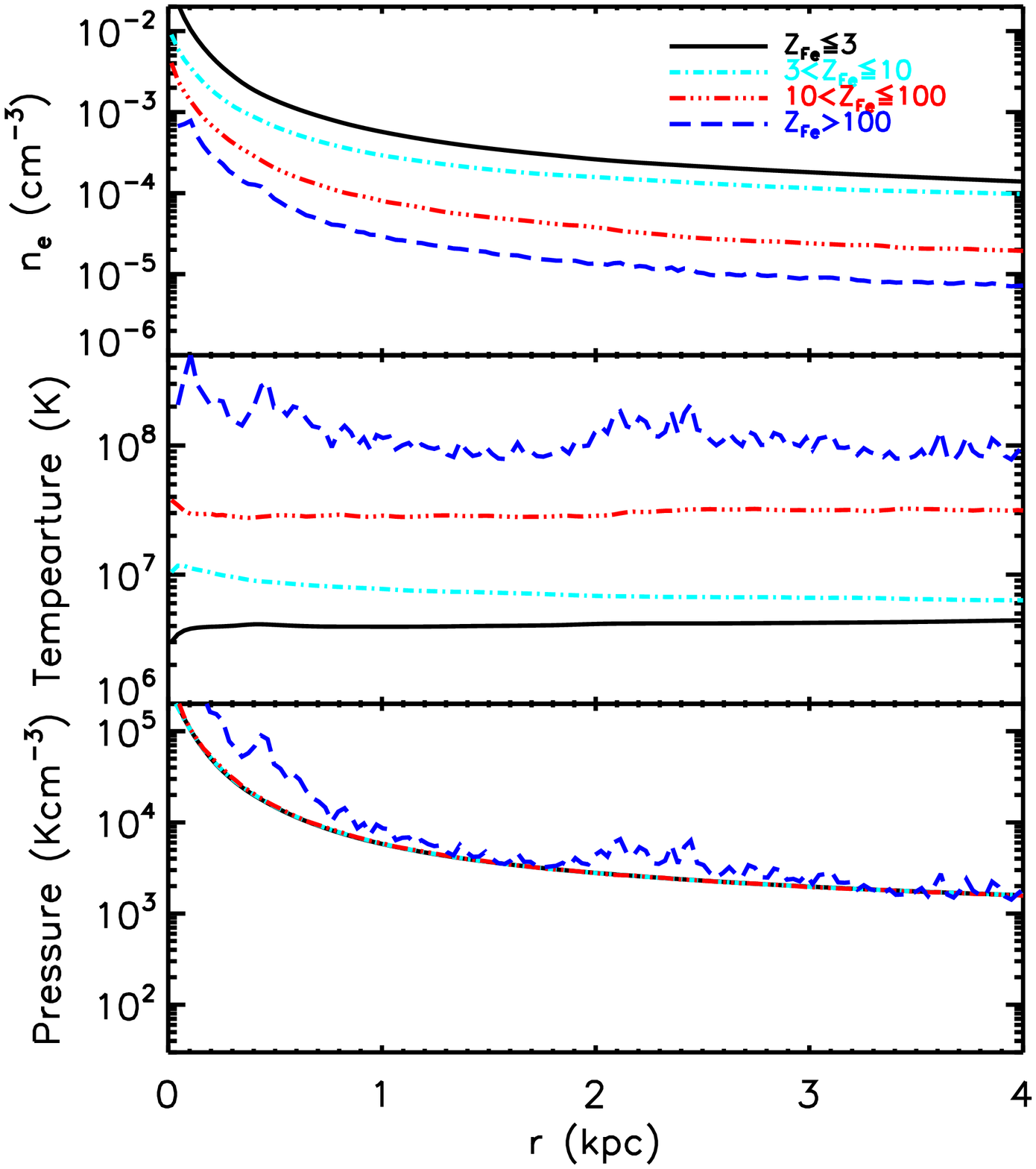}
\caption[The radial profiles of gas with
different $Z_{\rm Fe}$ in the supersonic outflow]
{\label{F:ZFeGas}
Average radial electron number density, temperature, and pressure
profiles for the supersonic (left panel) and subsonic
(right panel) cases. The profiles are plotted separately for the gas components 
with different iron abundances:
$Z_{\rm Fe} \leq 3 $ (solid black line),
$3 < Z_{\rm Fe} \leq 10$ (dash-dot cyan line),
$10 < Z_{\rm Fe} \leq 100$ (dash-three-dots red line), and
$Z_{\rm Fe} > 100$ (long-dash blue line).
}
\end{center}
\end{figure}

Fig.~\ref{F:ZFeGas} shows various averaged number density,
temperature, and pressure profiles extracted from the two simulations.
As expected, the profiles for the subsonic outflow
are much flatter than those for the supersonic ones.
The difference in the density, for example, is about one order of magnitude 
at large radii. The profiles of the hot gas components in four
iron abundance ranges show interesting
differences and similarities.
In general, the higher the abundance of a gas component is,
the smaller the number density or the higher the temperature is.
Except for the component with the highest abundance,
the other components all have similar pressure profiles.
This means that the bulk of the gas is approximately in a pressure equilibrium,
although the temperature and density distributions may still be 
highly structured. Similar phenomena are also reported for simulations of the SN
dominated interstellar medium in the Galaxy \citep{Avillez05,MacLow05,Joung06}
and in starburst regions \citep{Joung08}.

\epsscale{1.0}
\epsscale{1.0}

\begin{figure}[thbp]
\begin{center}
\plottwo{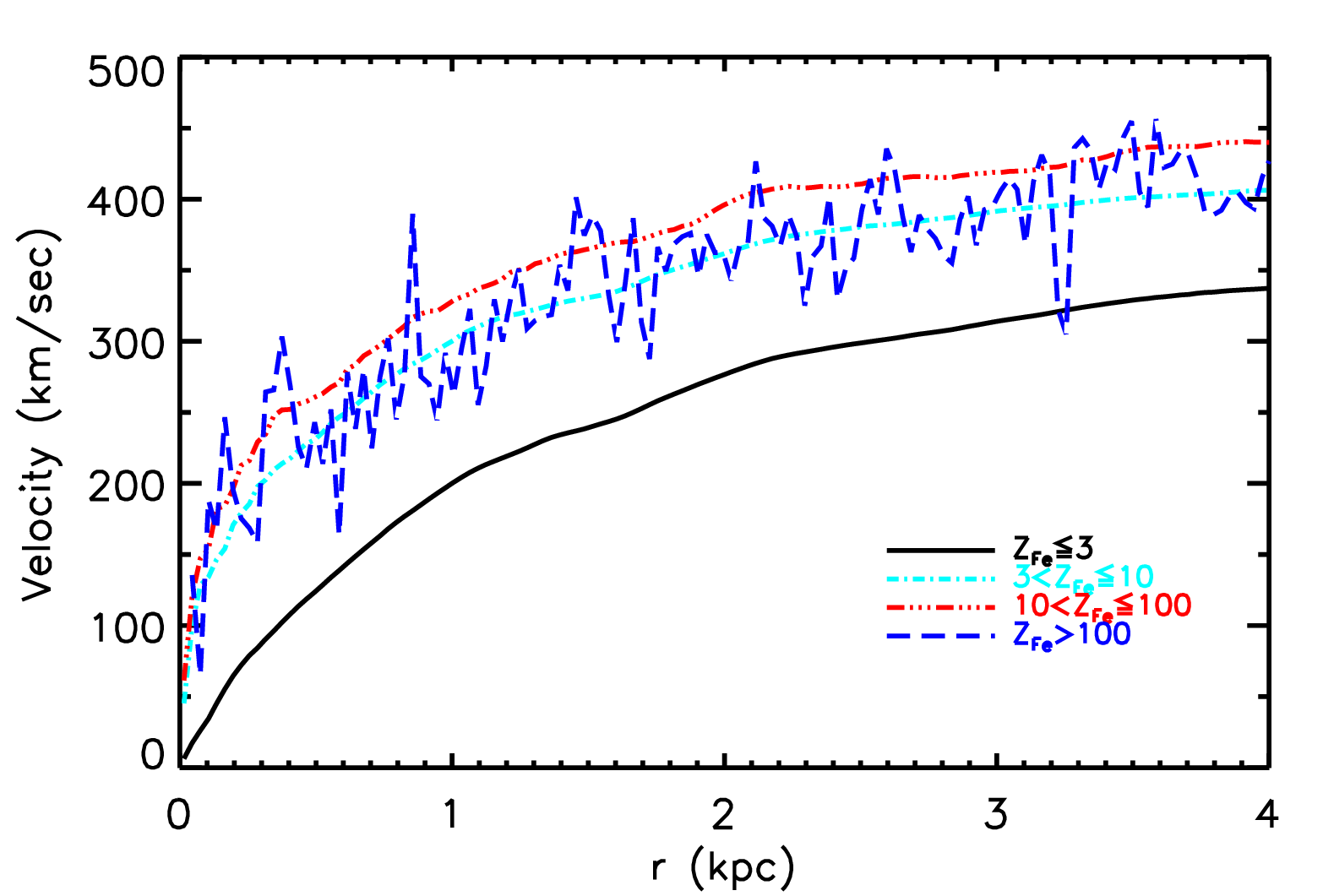}{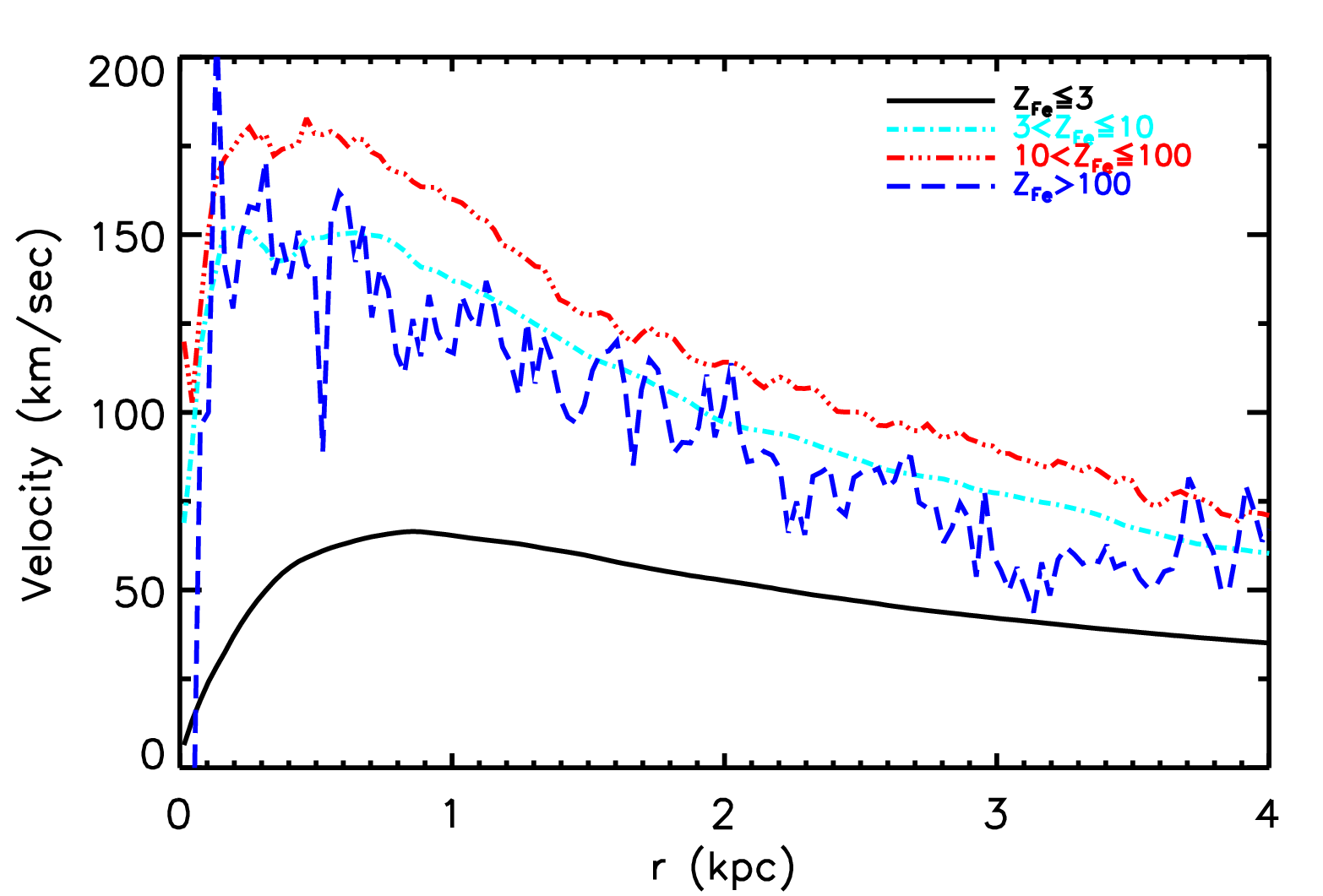}
\caption[The radial velocities of gas with different $Z_{\rm Fe}$]
{\label{F:ZFeGas_vr}
Average radial velocity profiles of the gas components with different
iron abundances for the supersonic (left) and subsonic (right) cases.
The line styles are the same as those in Fig.~\ref{F:ZFeGas}.
}
\end{center}
\end{figure}

Fig.~\ref{F:ZFeGas_vr} demonstrates that 
gas with higher iron abundance generally has a larger net
outward radial velocity.
The velocity of the gas component with $Z_{\rm Fe}>100$
fluctuates greatly, because it represents young SNRs
which may or may not have been fully accelerated by the buoyancy (see \S~4.2).
The gas component with  $10<Z_{\rm Fe} \leq 100$, representing
fully-accelerated and evolved SNRs with moderate dilution, 
has the largest velocity.

\epsscale{1.1}
\epsscale{1.0}

\begin{figure}[thbp]
\begin{center}
\plottwo{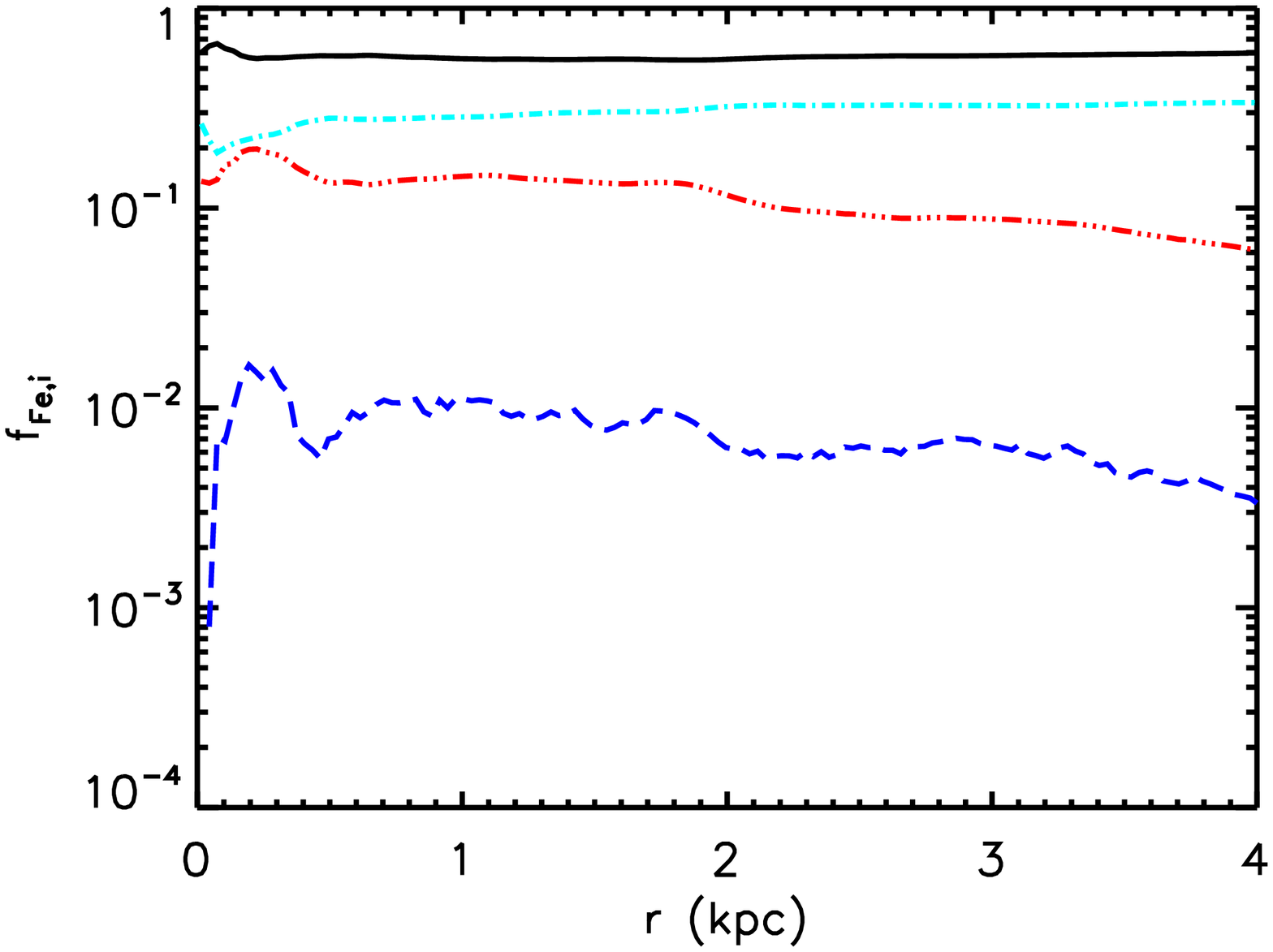}{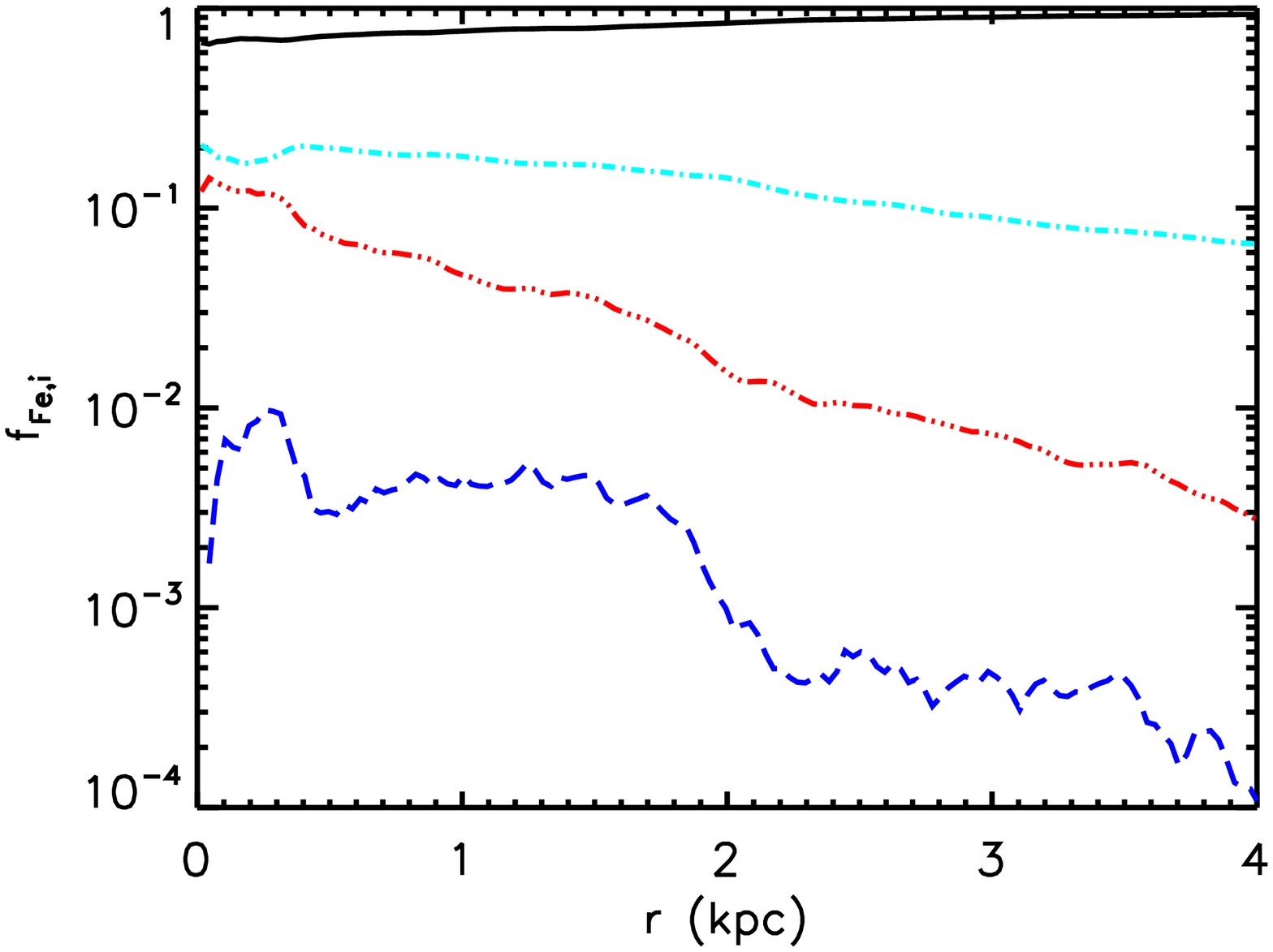}
\caption[The radial velocities of gas with different $Z_{\rm Fe}$]
{\label{F:ironfractions}
Average radial iron mass fractions of the individual gas components in the different abundance 
ranges for the supersonic (left) and subsonic (right) cases. The fractions
are relative to the total iron mass, which includes the iron mass from stars. 
The line styles are the same as those in Fig.~\ref{F:ZFeGas}.
}
\end{center}
\end{figure}

Fig.~\ref{F:ironfractions} shows the iron  mass fractions of 
the individual gas components in the different abundance ranges.
The fractions in iron-rich components decrease considerably faster with 
increasing radius in the subsonic case than in the supersonic one. This is because
the outflow velocity is much slower in the former case, leading to more 
dilution and mixing of the iron ejecta before moving to 
large galactic radii.

\epsscale{1.1}
\epsscale{1.0}

\begin{figure}[t]
\begin{center}
\plottwo{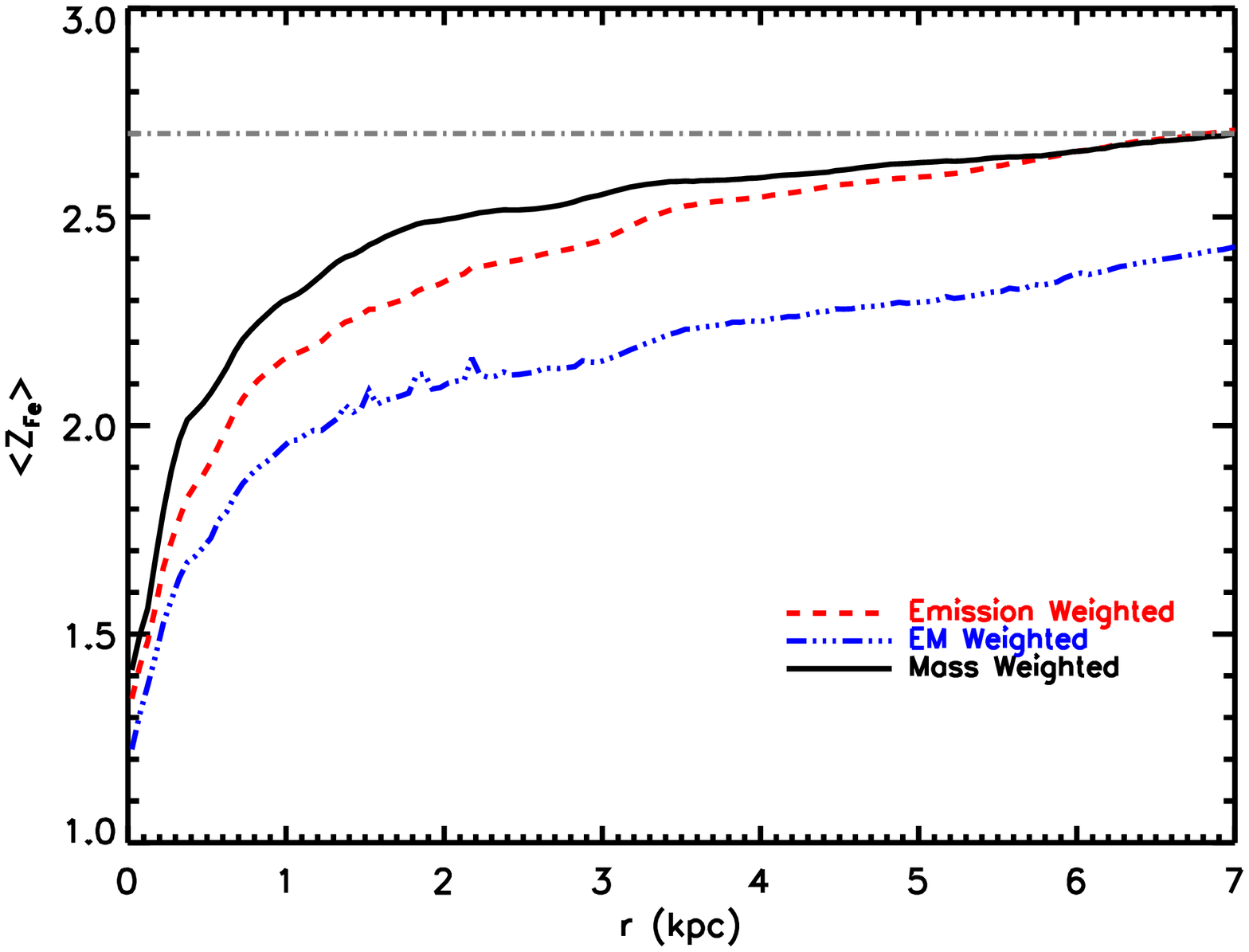}{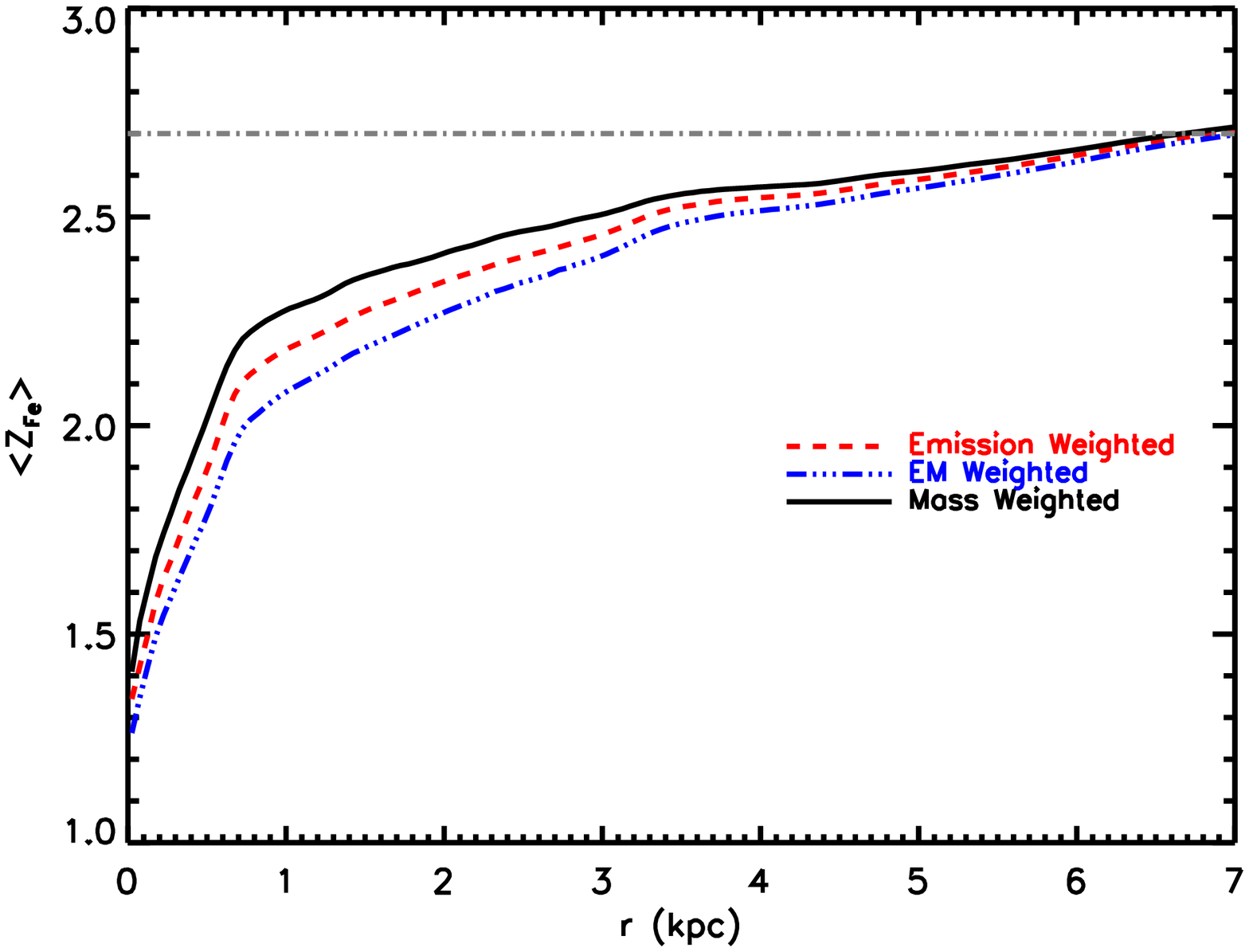}
\caption[The gradient of $Z_{\rm Fe}$]
{\label{F:ZFeGradient}
Average radial iron abundance profiles for the supersonic (left panel) and subsonic
(right panel) cases. The profiles weighted by mass, emission,
and emission measure are plotted separately as labeled.
The horizontal dash-dot gray line
denotes the expected abundance if iron ejecta is uniformly
mixed with stellar mass loss materials. 
}
\end{center}
\end{figure}

Fig.~\ref{F:ZFeGradient} presents the radial profiles of the averaged iron abundance,
weighted by three means. 
The mass-weighted iron abundance, defined as
\begin{equation}\label{eq:ZFebyM}
  <Z_{\rm Fe,mw}> = 
  \frac{\sum_i f_i \rho_i \Delta V_i}
  { \sum_i \rho_i \Delta V_i}
\end{equation}
where $f_i$ and $\rho_i$ is the \new{iron mass abundance } and mass density in a simulated 
sub-volume $\Delta V_i$, while the sum is over an interested region (e.g.,
a shell). $<Z_{\rm Fe,mw}>$ is thus directly proportional to the iron mass in
the region. Similarly, we can define the emission measure-weighted iron abundance
as
\begin{equation}\label{eq:ZFebyEM}
  <Z_{\rm Fe,emw}> = 
  \frac{\sum_i f_i  n_{e,i} n_{H,i}  \Delta V_i}
  { \sum_i n_{e,i} n_{H,i} \Delta V_i}, 
\end{equation}
where the electron number density $n_{e,i}$ is related to the hydrogen 
number density $n_{H,i}$ (hence $\rho_i$) 
and the $Z_{Fe,i}$ in the same way as described by
\citet{Brighenti2005}. $<Z_{\rm Fe,emw}>$  probably represents the 
simplest way to estimate the equivalent
abundance that can be compared with an actual measurement based on 
X-ray emission. An alternative model proxy to the X-ray emission-inferred 
value is the emission-weighted iron abundance, 
\begin{equation}\label{eq:ZFebyE}
  <Z_{\rm Fe,ew}> = 
  \frac{\sum_i f_i  n_{e,i} (n_{H,i}\Lambda_{\rm noFe}+
    n_{\rm _{Fe}} \Lambda_{\rm Fe}) 
    \Delta V_i}
  { \sum_i  n_{e,i} (n_{H,i}\Lambda_{\rm noFe}+
    n_{\rm _{Fe}} \Lambda_{\rm Fe}) \Delta V_i }.    
\end{equation}
In our simulations, the $<\!\!Z_{\rm Fe,ew}\!\!>$ might be overestimated
at large radii because SN iron cores are not adequately
resolved  (i.e., the spatial resolution in regions of $r>5$\,kpc
is reduced by a factor of four), which numerically
mixes the iron ejecta over a larger volume and
effectively results in a higher $n_e$ (hence stronger emission).
Fig.~\ref{F:ZFeGradient} shows that 
the radial trends of these differently weighted iron abundances are
qualitatively the same: a positive gradient with a minimum value at
the center! The lower limit of the abundance is
solar, assumed for the stellar mass loss. 

\epsscale{1.0}
\epsscale{1.0}

\section{Discussion}

Here we attempt to give a physical account of several key
phenomena observed in the simulations presented above.

\subsection{Thermal Evolution of Ia SN Iron Ejecta}

\begin{figure}[bpt]
\epsscale{1.1}
\begin{center}
\plottwo{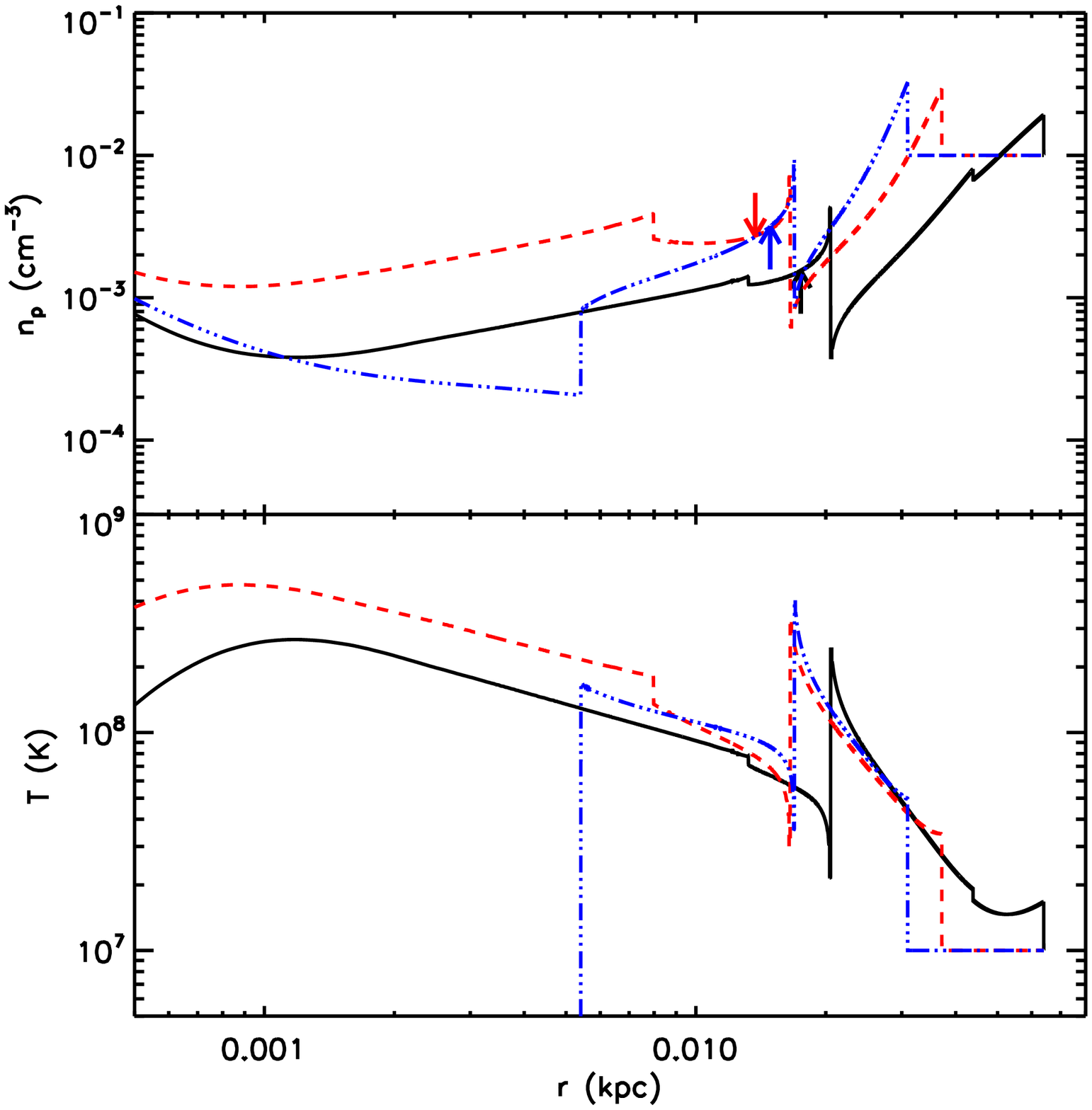}{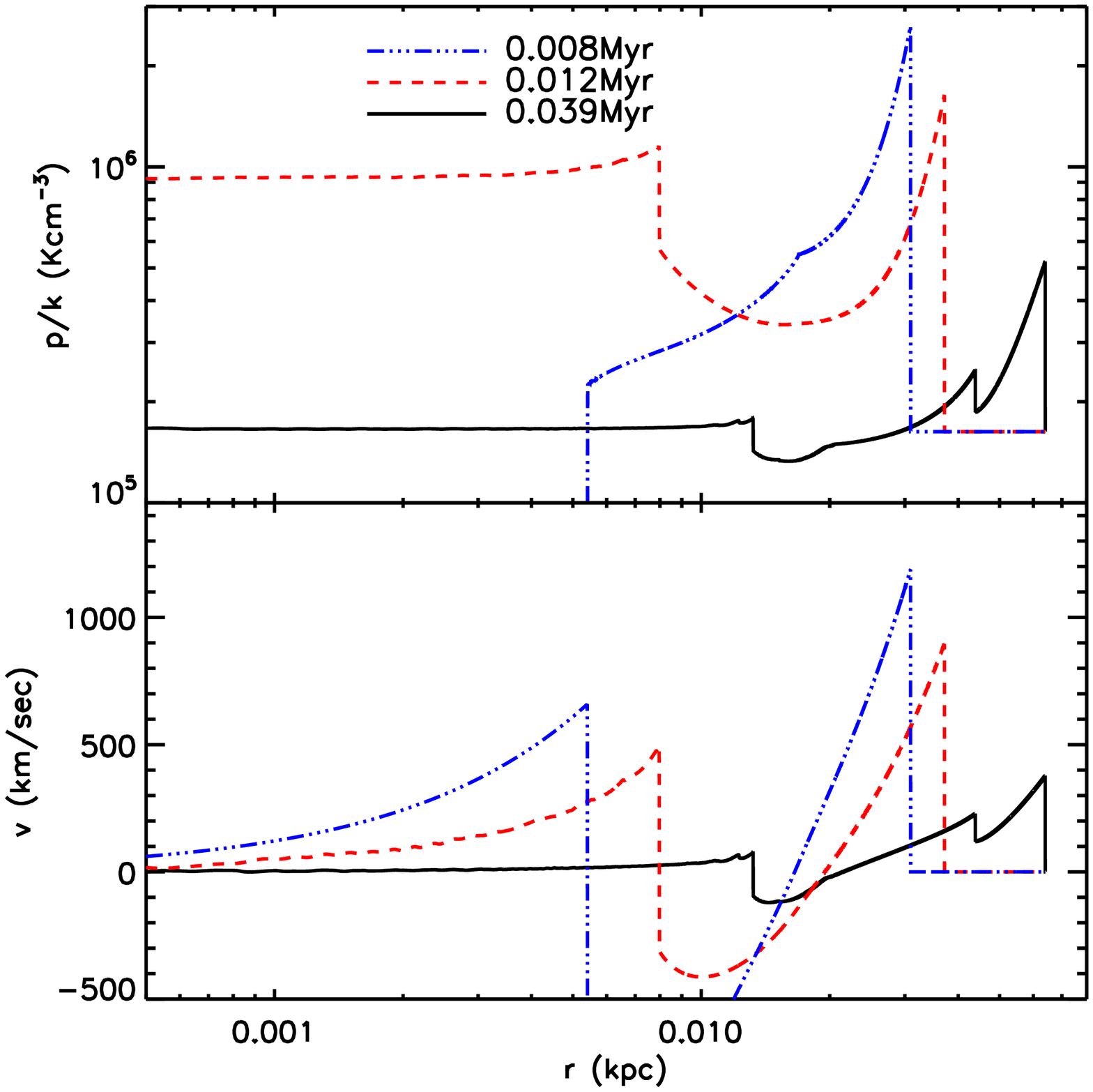}
\caption[The profiles are an SNR]{\label{F:SNejecta}
Sample density, temperature, pressure, and velocity profiles
of an SNR at three different
ages: $8\times 10^3\,$yr (dash-three-dots blue line),
$1.2\times 10^4\,$yr (dash red line),
$3.9\times 10^4\,$yr (solid black line).
The arrows in the density panel denote the
outer radii of the iron ejecta.
}
\end{center}
\epsscale{1.0}
\end{figure}

\begin{figure}[hbpt]
\epsscale{1.0}
\begin{center}
\plotone{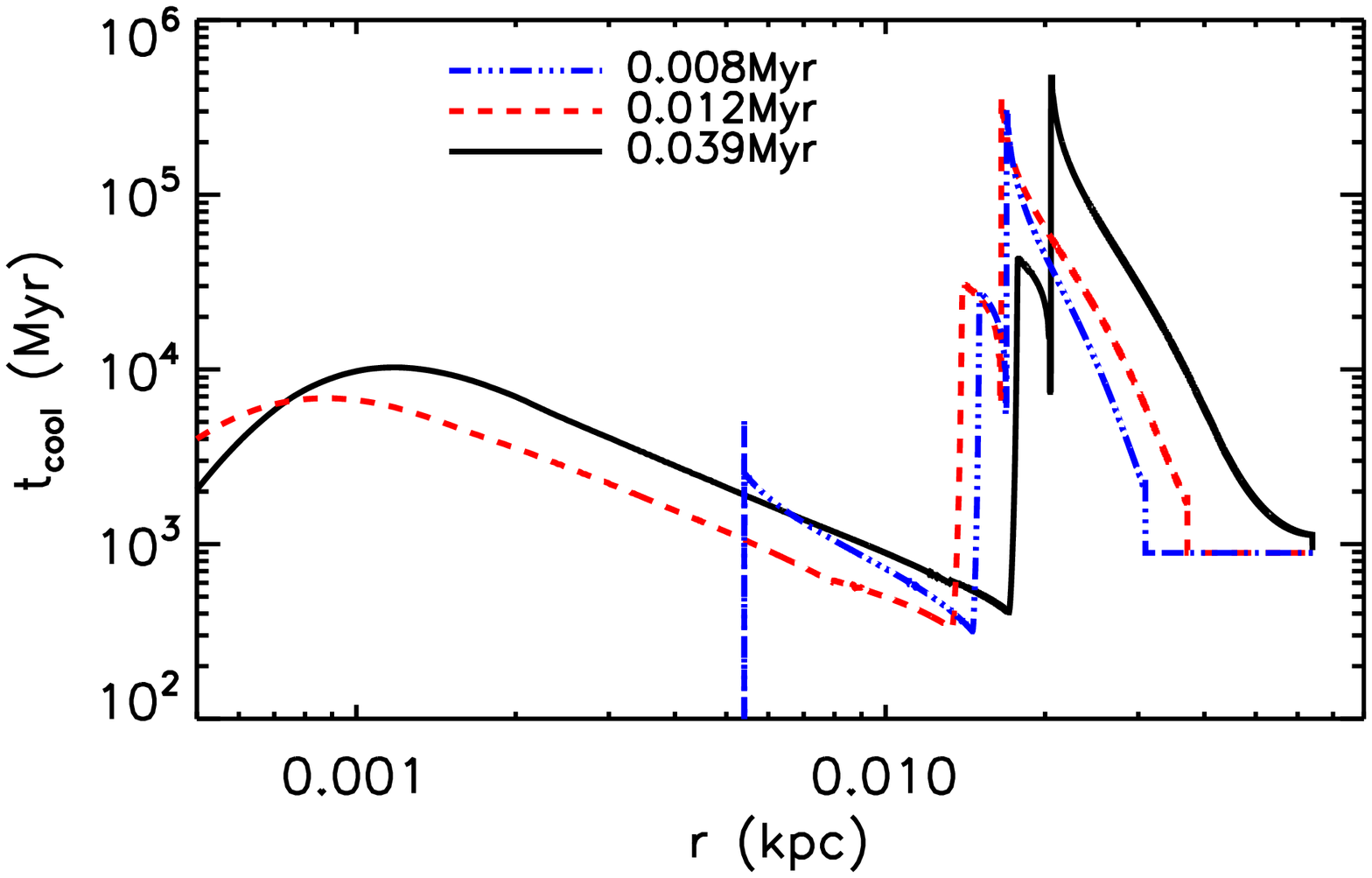}
\caption[The cooling time scale of Ia SN core]
{\label{F:snr_cooltime}
The cooling time scale as a function of radius, calculated with
the profiles given in Fig.~\ref{F:SNejecta}.
}
\end{center}
\epsscale{1.0}
\end{figure}

We find no evidence for significant cooling of the hot gas, especially the iron
ejecta, in our simulated region. To have a better understanding
of this issue, it is instructive to first consider a simple 1-D simulation
of a typical SNR in a hot gas environment, as in \citet{Brighenti2005}, but
in the context of an intermediate-mass spheroid considered here \citep{Tang05}.
As mentioned in \S~2, the 1-D simulated SNR is also used to 
generate the seeds for the 3-D simulations. Fig.~\ref{F:SNejecta} demonstrates
the simulated SNR evolution in a uniform ambient medium of temperature $T_0 = 10^7$\,K
and electron density $n_e = 0.01\, {\rm~cm^{-3}}$, appropriate
for a region near the center of the spheroid. The core that 
encloses 0.7\,$\rm M_\odot$ reaches a size of 15\,pc at the age of
$10^4$\, years and increases only slightly after that.
This is also roughly the time scale for the ejecta to be 
fully thermalized by the converging reverse shock. Then
the core  turns into a tenuous hot bubble,
with a very low density and high temperature. Outside the iron core is an
envelope of the rest of the SN ejecta (assumed to be 0.7M$_\odot$).
The swept-up material can also be heated to more
than $10^{8}$\,K only initially by the strong forward blastwave;
the averaged temperature of the swept-up material is considerably lower
because of the quick weakening of the expanding blastwave and the adiabatic 
cooling of the SNR. The radiative cooling time scale (Fig.~\ref{F:snr_cooltime})
for such an SNR is much too long to be important, compared to
the outflow time of the hot gas from a spheroid ($\lesssim 10^7$ yrs). 
In particular, the X-ray emission from the ejecta is largely undetectable. 

We now consider additional processes that an SNR
should experience in a spheroid.
One is the dilution due to mass injection from evolved stars
(or evaporation of cloud-lets). 
As the dilution takes place,  $n_e$ increases while 
$\Lambda_{_{\rm Fe}}(T)$ drops. The net result is that the cooling rate 
peaks when the iron abundance drops to $\sim 100$ solar 
\citep{Brighenti2005}. Further dilution tends to decrease the cooling rate.
Take the above 1-D SNR as an example.
It takes about 2.7\,Myr for the SNR core to accumulate an 
additional 5.4 \,$M_\odot$, which dilutes the ejecta to an iron abundance 
of 100 solar, the value to avoid rapid radiative cooling 
\citep{Brighenti2005}. In comparison, the time for the SNR
to flow this radius to 2\,kpc is about 5\,Myr.
Thus the dilution is important, although the core should still 
have a high iron abundance when being transported out of the 
spheroid. Another process is the mixing of the iron ejecta with the
ambient medium. The density of bulk of the SNR interior is always smaller than the 
surrounding medium, as demonstrated in the 1-D modeling,
and is likely subjected to the Rayleigh-Taylor instability
(e.g., \citealt{Sedov59,Wang01}).
The passage of blastwaves produced by nearby SNe also tends to accelerate
the mixing process, especially in the inner spheroid 
region where the stellar density is very high.
Our Euler-based hydrodynamical simulations account for this cooling rate
effect as well as the dilution and mixing processes. 

We conclude that the radiative cooling of the iron ejecta is unlikely
to be important in the type of spheroids we have considered. But we cannot 
rule out the possibility that Ia SN ejecta may cool and drop off from 
hot gas in a more quiescent environment. In a giant elliptical
galaxy considered by \citet{Brighenti2005}, hot gas density can be 
rather high, substantially increasing the cooling, and
the bulk motion may not be significant. Unfortunately, a 3-D simulation
of the hot gas in such an environment is still very difficult to conduct,
because of both the large dynamic range over a big volume and
the difficulty to handle the boundary condition.

\subsection{Buoyancy-Driven Dynamics of the Ejecta}

One of our findings is that the differential radial outflows play an
important role in determining the iron abundance profile of the hot gas.
The relatively large outflow speed of the iron-rich gas is apparently
caused by the presence of the buoyancy force. A 
reverse-shock heated SN ejecta bubble, with
a density that is one to two orders of magnitude less than that of 
the ambient medium,  is subject to a buoyancy-driven acceleration as 
\begin{equation}\label{eq:buoy}
  {\mathbf a}_{\rm buoy} = -{\mathbf g}\left(
    \frac{\bar{\rho}_{\rm amb}}{\bar{\rho}_{\rm hb}} - 1\right),
\end{equation}
where $\mathbf g$ is the gravitational acceleration while
$\bar{\rho}_{\rm hb}$ and $\bar{\rho}_{\rm amb}$ are the average densities of the
bubble and its ambient medium.
For our adopted model spheroid, $g\simeq 90\, \rm km\,s^{-1}\,Myr^{-1}$
at 0.5\,kpc, for example, and is larger near the
galactic center. Assuming a typical $\bar{\rho}_{\rm amb} / \bar{\rho}_{\rm hb} \approx 10$, the hot bubble can be accelerated over a period of 0.2\,Myr 
to about 150 ${\rm~km~s^{-1}}$, a typical velocity difference
between the iron-rich gas and the bulk of the outflow. This
velocity is also about the equilibrium value 
that can be reached when the buoyancy force is balanced by a drag:
\begin{equation}\label{eq:drag}
  {\mathbf a}_{\rm drag} = - \hat{\mathbf v}_{\rm hb}
  \cdot \frac{1}{2} C A_{\rm hb} v_{\rm hb}^2
  \rho_{\rm amb} / m_{\rm hb}, 
\end{equation}
where $A_{\rm hb}$, $m_{\rm hb}$, and $\hat{\mathbf v}_{\rm hb}$
are the maximum cross-section area, mass, and velocity
of the hot bubble, respectively. The coefficient $C$
is suggested to be $\sim 1$ (e.g., \citealt{Jones96}).
The above crude estimates are consistent with the results from
our detailed examination of 
individual hot bubbles in the simulations, which naturally account for 
such complications as the changing sizes and shapes as well as
the decreasing density contrasts relative to the ambient medium.

\subsection{Nature of the Iron Abundance Gradient}

We find that the positive radial gradient of the iron abundance shown in
Fig.~\ref{F:ZFeGradient} is due largely to
the differential outward velocities of the gas (Fig.~\ref{F:ZFeGas_vr}).
The relative fast outward motion of the iron-rich gas naturally
leads to a reduced iron mass fraction in the bulk outflow.
For an illustration, suppose that a model hot gas contains only two components: 
the first one (sub-scripted with 1 hereafter) is iron-free,
whereas the second one (sub-scripted with 2) is pure iron. These two components
have outward mass fluxes at radius $r$:
\begin{eqnarray}\label{eq:fluxconv}
  F_1 = 4\pi r^2 \rho f_1 v_1, \\
  F_2 = 4\pi r^2 \rho f_2 v_2,
\end{eqnarray}
where $f$ and $v$ represent the average mass fraction and outward velocity
of each component, while $\rho$ is the total mass density of the gas.
The iron abundance is then 
\begin{eqnarray}
  Z_{\rm Fe} & \equiv & \frac{f_2}{f_1 + f_2} \\
  & \simeq &\frac{F_2}{F_1} \frac{v_1}{v_2},
  \label{eq:Feabundef}  
\end{eqnarray}
where $f_2 << f_1$ is adopted because the total iron mass is no more than
a few percent of the total gas mass.
The ratio $F_2$/$F_1$ is the same as the
iron mass fraction of the injection from the stellar mass loss,
independent of the spatial lumpiness in the iron distribution.
Eq.~\ref{eq:Feabundef} thus illustrates that the faster the iron-rich gas component
moves relative to the rest of the outflow,
the lower its mean iron abundance should be.

\section{Summary and Conclusions}

We have examined the evolution of Ia SN iron ejecta
in the hot gas undergoing outflows from a typical intermediate-mass 
galactic stellar spheroid. 3-D hydrodynamic simulations of this hot gas
are conducted for both supersonic and subsonic cases.
These simulations show no evidence for the iron ejecta to 
significantly cool and drop out of the hot gas. 
The largely inhomogeneous enrichment and heating, inherited from sporadic Ia 
SNe, can significantly affect the X-ray measurement of the mean iron abundance 
of the hot gas in two ways:

\begin{enumerate} 
\item Hot and low-density iron bubbles, generated by Ia SNe, tend to move 
outward substantially faster than the ambient medium. 
This differential motion, produced mainly 
in the inner galactic region under the strong buoyancy force, effectively reduces
the mean iron abundance of the hot gas.
The reduction decreases with the increasing galactic radius as
the iron bubbles are gradually mixed with the ambient medium and are
diluted by fresh mass injection from evolved stars.

\item Because of their high temperature and low density, the iron-rich 
bubbles hardly radiate, contributing little to the X-ray emission
of the hot gas \citep{Tang09a}. The bulk of the emission
arises from the ambient medium swept up by SNR forward blastwaves, which
has higher densities and lower temperatures. This
iron segregation as well as the density and temperature distributions, 
most apparent again in the inner galactic region,
can lead to a significant underestimate of the actual
mean iron abundance of the hot gas, when characterized with a simplistic 
plasma model (e.g., with one- or two-temperature components).
\end{enumerate}

These two effects together provide a natural explanation for
the low iron abundance and its positive radial gradient, 
as inferred from existing X-ray measurements of hot gas in 
various galactic stellar spheroids. 
\new{Unfortunately, a direct comparison with existing observations is still 
difficult. The inward drop of the iron abundance is usually observed in 
massive spheroids. However, this is most likely an observational bias. We are
not aware of a reliable measurement of the abundance profile for low- and
intermediate-mass spheroids. Not only the counting statistics is typically
insufficient, but the contribution from faint
cataclysmic variables and coronally active binaries becomes important
as well. This contribution, in particular, is typically
not subtracted in existing studies. A more careful data analysis is needed
for such spheroids.
The observational measurements for massive objects are easier to make
and more reliable. The measurement by Buote et al. (2003), 
for example, shows that the iron abundance within the central
$\sim 4$ kpc radius of NGC 5044 is on average only $\sim 2/3$ of
that in the 4-12 kpc surrounding region, which is consistent with the 
prediction from our simulation of the subsonic outflow. 
However, NGC 5044 is a massive galaxy in a group, while the simulation is 
suited for an intermediate-mass one. Thus,
a quantitative comparison would be problematic, although the 3-D effect of 
the buoyancy on the abundance distribution should remain the same
qualitatively.}

This work together with those presented in \citet{Tang09a,Tang09b} demonstrates
that X-ray observations of hot gas in and around spheroids can potentially
be used as a unique tool to probe the dynamics of the 
feedback from evolved stars as well as SMBHs, which represents a very poorly
understood part of galaxy formation and evolution theories.
To reach this goal, more work is clearly needed to understand the hot gas,
both theoretically and observationally. Particularly useful will be 
spatially-resolved 
high-resolution spectroscopy of the hot gas, as could be provided by the 
future International X-ray Observatory, as well as careful analysis 
of existing high-quality X-ray imaging data. But a large parameter space of the 
modeling also needs to be explored to facilitate quantitative comparisons with 
the existing and future observations.

\acknowledgments
We thank the referee for useful comments.
The software used in this work was in part developed
by the DOE-supported ASC/Alliance Center for Astrophysical
Thermonuclear Flashes at the University of Chicago.
Simulations were performed at the Pittsburgh Supercomputing Center
supported by the NSF.
We also acknowledge the support by NASA through grants NNX06AI18G and TM7-8005X (via
SAO/CXC).


\begin{thebibliography}{}
\setlength{\itemsep}{-0.25truecm}
\def\aap{A\&A}
\def\apj{ApJ}
\def\apjs{ApJS}
\def\mnras{MNRAS}
\def\araa{ARA\&A}
\def\nat{Nature}

\bibitem[Anders \& Grevesse (1989)]{Anders89}
Anders E., \& Grevesse N., 1989, Geochimica et Cosmochimica Acta, 53, 197
\bibitem[Borgani et~al. (2008)]{Borgani08} Borgani S., Fabian D., Tornatore L., Schindler S., Dolag K., Diaferio A., 2008, Space Sci. Review, 134, 379
\bibitem[Brighenti \& Mathews (2005)]{Brighenti2005}
Brighenti Fabrizio, \& Mathews William G., 2005, ApJ, 630, 864
\bibitem[Buote (2000a)]{Buote00apj}
Buote D. A., 2000a, ApJ, 539, 172
\bibitem[Buote (2000b)]{Buote00}
Buote D. A., 2000b, MNRAS, 311, 176
\bibitem[Buote et al. (2003)]{buo03}
Buote D. A., Lewis A. D., Brighenti F., \& Mathews W. G. 2003, ApJ, 595, 151
\bibitem[Cioffi \etal (1988)]{Cioffi1988}
Cioffi, D. F., Mckee, C. F., \& Bertschinger, E. 1988, ApJ, 334, 252
\bibitem[David et~al. (2006)]{David06}
David L. P., Jones C., Forman W., Vargas I. M., Nulsen P., 2006, ApJ, 653, 207
\bibitem[de Avillez \& Breitschwerdt (2005)]{Avillez05}
de Avillez M. A., \& Breitschwerdt D., 2005, A\&A, 436, 585
\bibitem[Diehl \& Statler (2008)]{die08} Diehl, S, \& Statler, T. S. 2008, ApJ, 680, 897
\bibitem[Fabjan et~al. (2008)]{Fabjan08}
Fabian D., Tornatore L., Borgani S., Saro A., Dolag K., 2008, MNRAS, 386, 1265
\bibitem[Fryxell \etal (2000)]{Fryxell00}
Fryxell B., et al., 2000, APJS, 131, 273
\bibitem[Gastaldello \& Molendi (2002)]{gm02}
Gastaldello F., \& Molendi S. 2002, ApJ, 572, 160
\bibitem[Hernquist (1990)]{Hernquist1990}
Hernquist L., 1990, ApJ, 356, 359
\bibitem[Humphrey \& Buote (2006)]{Humphrey06}
Humphrey P. J., \& Buote D. A., 2006, ApJ, 639, 136
\bibitem[Ji et al. (2009)]{jij09} Ji J., Irwin J. A., Athey A., Bregman J. N., \& Lloyd-Davies E. J. 2009, ApJ, 696, 2252
\bibitem[Jones et~al. (1996)]{Jones96}
Jones T. W., Ryu D., \& Tregillis I. L., 1996, ApJ, 473, 365
\bibitem[Joung et~al. (2008)]{Joung08}
Joung M. K., Mac Low M.-M., Bryan G. L., 2008, ApJ, 704, 137
\bibitem[Joung \& Mac Low (2006)]{Joung06}
Joung M. K., \& Mac Low M.-M., 2006, ApJ, 653, 1266
\bibitem[Liedahl et~al. (1995)]{Liedahl95}
Liedahl D. A., Osterheld A. L., \& Goldstein W. H., 1995, ApJ, 438, L115
\bibitem[Lowenstein \& Mathews (1987)]{LM1987}
Loewenstein M., Mathews W. G., 1987, ApJ, 319, 614
\bibitem[Mac Low et~al. (2005)]{MacLow05}
Mac Low M.-M., Balsara D. S., Kim J., de Avillez M. A., 2005, ApJ, 626, 864
\bibitem[Mathews \& Brighenti (2003)]{Mathews2003}
Mathews, W. G., \& Brighenti, F. 2003, ARA\&A, 41, 191
\bibitem[Mewe et~al  (1985)]{Mewe85}
Mewe R., Gronenschild E. H. B. M., \& van den Oord G. H. J. 1985, A\&AS, 62, 197
\bibitem[Raley et~al. (2007)]{Raley07}
Raley E. A., Shelton R. L., Plewa T., 2007, ApJ, 661, 222
\bibitem[Sedov (1959)]{Sedov59}
Sedov L. I 1959, Similarity and Dimensional Methods in Mechanics, translation from 4th Russian edition, Academic press New York and London
\bibitem[Strickland \& Stevens (1998)]{Strickland98}
Strickland D. K., \& Stevens I. R., 1998, MNRAS, 297, 747
\bibitem[Sutherland \& Dopita (1993)]{Sutherland93}
Sutherland R. S., Dopita M. A., 1993, ApJS, 88, 253
\bibitem[Tang \& Wang (2005)]{Tang05}
Tang S., \& Wang Q. D. 2005, ApJ, 628, 205
\bibitem[Tang \& Wang (2009)]{tw09}
Tang, S, \& Wang, Q. D. 2009, MNRAS, 397, 2106
\bibitem[Tang et~al. (2009a)]{Tang09a}
Tang S., Wang Q. D., Lu Y., Mo H. J., 2009a, MNRAS, 392, 77
\bibitem[Tang et~al. (2009b)]{Tang09b}
Tang S., Wang Q. D., Mac Low M.-M., Joung M. R. 2009b, MNRAS, 398, 1468
\bibitem[Trinchieri et~al. (2008)]{tri08}
Trinchieri, G., et al. 2008, ApJ, 688, 1000
\bibitem[Wang \& Chevalier (2001)]{Wang01}
Wang C., \& Chevalier R. A., 2001, ApJ, 549, 1119
\end{thebibliography}
\end{document}